# Highly sensitive electromechanical piezoresistive pressure sensors based on large-area layered PtSe$_2$ films

*Stefan Wagner[1], Chanyoung Yim[2], Niall McEvoy[3], Satender Kataria[1], Volkan Yokaribas[4], Agnieszka Kuc[5], Stephan Pindl[6], Claus-Peter Fritzen[4], Thomas Heine[5], Georg S. Duesberg[3], Max C. Lemme[1,7*]*

[1]Chair of Electronic Devices, Faculty of Electrical Engineering and Information Technology, RWTH Aachen University, Otto-Blumenthal-Str. 2, 52074 Aachen, Germany

[2]Institute of Physics, EIT 2, Faculty of Electrical Engineering and Information Technology, Universität der Bundeswehr München, Werner-Heisenberg-Weg 39, 85577 Neubiberg, Germany

[3]Centre for Research on Adaptive Nanostructures and Nanodevices (CRANN), Advanced Materials and BioEngineering Research (AMBER) and School of Chemistry, Trinity College Dublin, Dublin 2, Ireland

[4]Department of Mechanical Engineering, University of Siegen, 57076 Siegen, Germany

[5]Wilhelm-Ostwald-Institute für Physikalische und Theoretische Chemie, Universität Leipzig, Linné Str. 2, 04103 Leipzig, Germany

[6]Infineon Technologies AG, Wernerwerkstraße 2, 93049 Regensburg, Germany

[7]Advanced Microelectronic Center Aachen (AMICA), AMO GmbH, Otto-Blumenthal-Str. 25, 52074 Aachen, Germany

ABSTRACT: Two-dimensional (2D) layered materials are ideal for micro- and nanoelectromechanical systems (MEMS/NEMS) due to their ultimate thinness. Platinum diselenide ($PtSe_2$), an exciting and unexplored 2D transition metal dichalcogenides (TMD) material, is particularly interesting because its scalable and low temperature growth process is compatible with silicon technology. Here, we explore the potential of thin $PtSe_2$ films as electromechanical piezoresistive sensors. All experiments have been conducted with semi-metallic $PtSe_2$ films grown by thermally assisted conversion of Pt at a CMOS-compatible temperature of 400°C. We report high negative gauge factors of up to -84.8 obtained experimentally from $PtSe_2$ strain gauges in a bending cantilever beam setup. Integrated NEMS piezoresistive pressure sensors with freestanding PMMA/$PtSe_2$ membranes confirm the negative gauge factor and exhibit very high sensitivity, outperforming previously reported values by orders of magnitude. We employ density functional theory (DFT) calculations to understand the origin of the measured negative gauge factor. Our results suggest $PtSe_2$ as a very promising candidate for future NEMS applications, including integration into CMOS production lines.

Layered two-dimensional (2D) materials have extraordinary electrical, optical and mechanical properties that suggest high potential for a wide range of nanoelectronics applications.[1,2] 2D transition metal dichalcogenides (TMDs) have recently been intensively investigated due to their wide range of inherent electronic properties that complement graphene. Out of this family, platinum diselenide ($PtSe_2$) is a relatively new and thus far little explored TMD material. Monolayer $PtSe_2$ is a semiconductor with a band gap of 1.2 eV,[3] while bulk $PtSe_2$ becomes semi-metallic with zero band gap.[4,5] It has been grown epitaxially[6] or synthesized on insulating substrates using thermally assisted conversion (TAC) of predeposited platinum (Pt) films by a vaporized solid selenium precursor.[4,7] The latter process can be carried out at temperatures between 400 °C and 450 °C, which is compatible with standard semiconductor back end of line (BEOL) processing,[8] in contrast to other 2D materials.[9,10] We have demonstrated potential applications of TAC $PtSe_2$ in highly sensitive gas sensors and wide spectral photodetectors with equal or even superior performance compared to graphene and other TMD-based devices.[7,11,12] Here, we explore the electromechanical properties of $PtSe_2$ in nanoelectromechanical (NEMS) pressure sensors. The devices utilize the piezoresistive effect in $PtSe_2$, which gives rise to a change in its electrical resistance upon mechanical strain. We investigate $PtSe_2$-membrane based pressure sensors and find sensitivities that are orders of magnitude higher than those of nanomaterial-based devices[13–16], including graphene[17–19]. Furthermore, we obtain an average negative gauge factor (GF) of -84.8 for $PtSe_2$ from a bending beam setup. Detailed density functional theory (DFT) simulations reveal an increase in the density of states (DOS) at the Fermi level of bulk $PtSe_2$, corroborating the experimentally observed piezoresistive properties.

Polycrystalline $PtSe_2$ films with nanometer sized grains were grown on silicon/silicon dioxide ($Si/SiO_2$) substrates by TAC at 400 °C of predeposited 1 nm thick Pt films (Figure S1 in the supporting information). Comprehensive characterization of layered PtSe$_2$ thin films synthesized in this manner, including X-ray photoelectron spectroscopy and scanning transmission electron microscopy is presented in our previous work.[4,7,11] The thickness and roughness of the layers was determined by atomic force microscopy (AFM, Figure 1a,b). A thickness of 4.5 nm was found for the selenized 1 nm initial Pt film. Assuming an interlayer distance of 0.6 nm for PtSe2[20], this corresponds to 7-8 layers for the 4.5 nm thick PtSe$_2$ film. The root mean square (RMS) roughness of the layer was 0.284 nm. The Raman spectrum of PtSe$_2$ film shows two sharp and prominent peaks corresponding to the E$_g$ and A$_{1g}$ modes of the 1T phase, attesting the crystalline nature of the films [4] (Figure 1c). After the film growth, a polymethyl metacrylate (PMMA) was spin-coated on top of the PtSe$_2$.

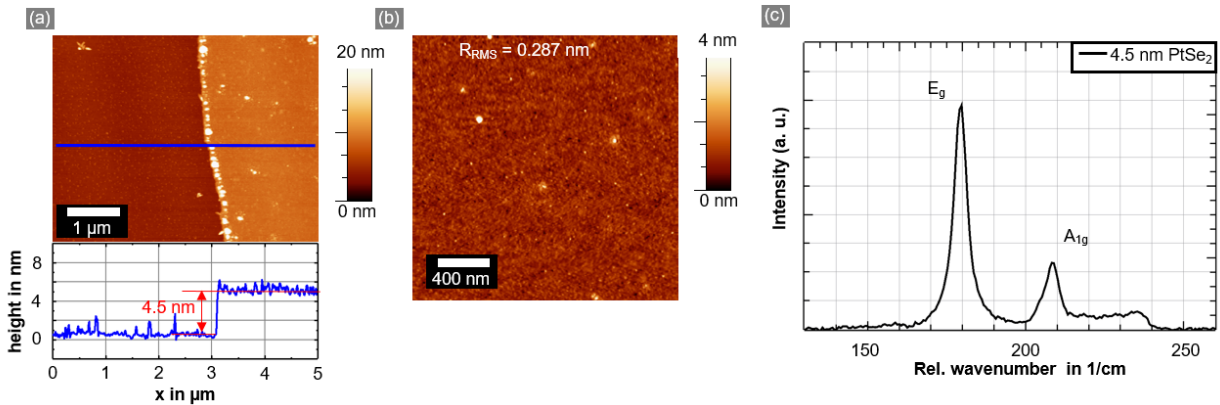

**Figure 1.** As-grown PtSe$_2$ film characterization: (a) AFM measurement of the selenized 1 nm initial Pt film including height profile of the step between the Si/SiO$_2$ substrate to the PtSe$_2$ film; (b) High-resolution AFM measurement displaying the RMS roughness; (c) Raman spectrum of the as-grown 4.5 nm thick PtSe$_2$ sample.

These centimeter-scale PtSe$_2$ thin films were transferred, using a typical polymer supported transfer process as described for graphene[21], onto the substrate comprising of 79 independent pressure devices. Each device consists of cavities covered by a patterned PtSe$_2$ film located in between metal contacts (Figure 2a-c). The PMMA layer was not removed after the transfer in order to stabilize and protect the PtSe$_2$ layer against environmental influences and to reduce possible cross sensitivity of the sensor. Details of the fabrication processes are described in the Experimental section and in the supporting information Figure S2a-d. A non-invasive Raman tomography cross-section, as demonstrated recently for graphene membranes,[22] confirms the presence of free-standing PtSe$_2$ across the cavity (Figure 2d,e) where devices with suspended PtSe$_2$ (Figure 2e, top) and collapsed membrane (Figure 2e, bottom) were measured and can clearly be distinguished with this method. The corresponding spectrum of the suspended PtSe$_2$ shows the characteristic E$_g$ and A$_{1g}$ modes without the Si(I) peak (Figure 2d, top), whereas the spectrum taken at the bottom of the trench shows only the characteristic Si(I) peak of the Si substrate (Figure 2d, bottom). This clear separation is feasible due to the 1.4 $\mu$m deep cavity, which defines the distance between the suspended PtSe$_2$ and the Si substrate.

The operation of the investigated pressure sensors can be described in the following way: the PtSe$_2$/PMMA membranes seal the cavities and trap environmental pressure inside them.[23-25] A different pressure (lower or higher) outside the cavity leads to deflection of the PtSe$_2$ membrane. This deflection induces strain in the PtSe$_2$ layer, resulting in a change of its resistance. The chip design used in the present experiments includes devices of similar size without cavities as control references. The pressure sensors, after wire-bonding and packaging into a chip socket, were placed in a home-built pressure chamber where pressure can be regulated from 1000 mbar (environmental pressure) down to 200 mbar (supporting information Figure S3). The pressure, humidity and temperature inside the chamber are recorded and controlled with commercial sensors during the experiments. Argon (Ar) atmosphere was used to exclude cross-sensitivity effects with gases and/or humidity.[7] The current-voltage (I-V) characteristics of the PtSe$_2$ device for environmental pressure shows good linearity, i.e. the transferred PtSe$_2$ forms good Ohmic contacts with metal (copper) electrodes (supporting information Figure S4). The measured data is presented in Figure 2f along with the simultaneously measured pressure (dashed red line, right y-axis). The resistance across the PtSe$_2$ sensors decreases with applied increasing strain on the membrane (caused by the decrease in pressure, supporting information Figure S5a). A pressure difference of 800 mbar between the outside and the inside of the cavity leads roughly to a 7% of change in resistance.

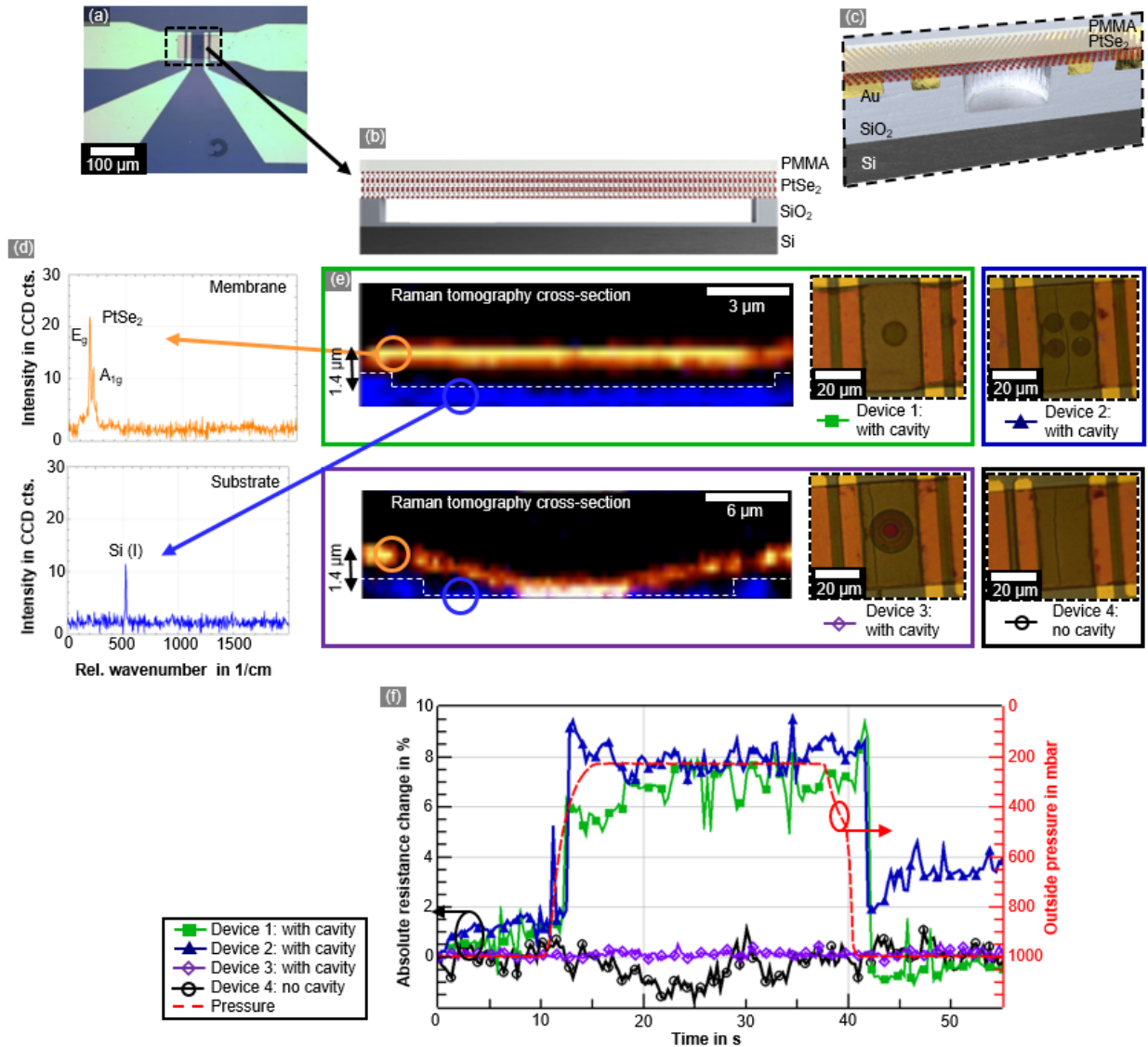

**Figure 2.** Pressure sensor fabrication and measurement. (a) Optical micrograph of one device with Au contacts and the PtSe$_2$ channel across the cavity area. (b) Schematic cross-section of the cavity area with suspended PtSe$_2$. (c) Schematic of one device including contacts and cavity area. (d) Raman point spectrum of the PtSe$_2$ membrane area and the Si substrate. (e) Raman tomography cross-sections of one device with suspended PtSe$_2$ (device 1) and one with collapsed

membrane (device 3). On the right are the corresponding optical micrographs of the cavity areas (device 1 and 3) as well as one other working device (device 2) and one device without cavity (device 4) as reference.(f) Resistance change in % for one pressure cycle (vented, pumped down to 200 mbar, vented) of the four devices shown in (e) (left y-axis) and the outside pressure (right y-axis).

The results of four devices are displayed in Figure 2f: two intact membranes on cavities (filled green squares and filled blue triangles), one with a collapsed membrane (empty purple diamond) and one for a reference device without cavity (empty black circles). A slight decrease (increase) in humidity during evacuation (venting) of the chamber is commonly observed for all measurements (supporting information Figure S5b), but does not give rise to any change in the resistance. Cross-sensitivity towards humidity can thus be excluded from having any substantial influence on the device behavior. We conclude that the observed resistance change is primarily caused by pressure (i.e. strain) changes during the measurements.

The absolute sensitivity ($S$) of a piezoresistive device is calculated using equation (1), where $\Delta R/R$ is the absolute value of the relative resistance change and $\Delta P$ is the pressure change during the sensor operation.[17]

$$S = \frac{\Delta R}{R \cdot \Delta P} \quad (1)$$

A total of 19 working devices were measured on four different chips through 69 different measurements (Figure 3a). A sensitivity of $1.05 \times 10^{-4}$ mbar$^{-1}$ was extracted from Figure 2f. An average of $5.51 \times 10^{-4}$ mbar$^{-1}$ was calculated for all measured devices (best device not included). The results were compared to some other NEMS pressure sensors which operate in a similar pressure range (Figure 3b,c). These use gallium arsenide (GaAs)[26], Si nanowires (Si NWs)[14,15], single- and multi-wall carbon nanotubes (SWNT and MWNT)[13,16] and graphene[17-19] as the piezoresistive materials. However, for most of these (except Smith et al.[17] and Fung et al.[16]) a separate membrane is needed which is treated, or where the piezoresistive materials are grown or applied, in a separate fabrication step. This adds additional complexity to the fabrication process, which is not the case for the PtSe$_2$-based pressure sensors, because the material acts as membrane and sensor at the same time. The thickness of other material is limited due to their mechanical stability. The measured sensitivity outperforms the best reported 2D and low dimensional materials based pressure sensors by a factor between 3 and 70 (Figure 3b). The SWNT-based pressure sensor[13] shows an equal sensitivity, however the membrane areas are nearly two to four orders of magnitude larger than the PtSe$_2$ membrane. To take this into account the sensitivity can be normalized to the membrane area to receive a direct comparison between the devices independent of their membrane size[27,17]. Here, the PtSe$_2$-based pressure sensor shows nearly 2 to 5 orders of magnitude higher sensitivity than the other devices (Figure 3c, see also supporting information Table S6).

We further point out that one of the devices showed an exceptionally high sensitivity of $1.64 \times 10^{-3}$ mbar$^{-1}$ (Figure S5c). Even though it could not be reproduced in other devices, this value was repeatedly measured for that particular device (Figure 3a, Chip 2, device 1). The sensitivity and normalized sensitivity of this device are plotted in Figure 3b,c (black star). When normalized to the membrane area, this device is over 3 orders of magnitude more sensitive than the highest performing devices shown in Figure 3c (a SWNT[13] and a graphene membrane[17] based device), and 4 to 6 orders of magnitude more sensitive than other reported devices.

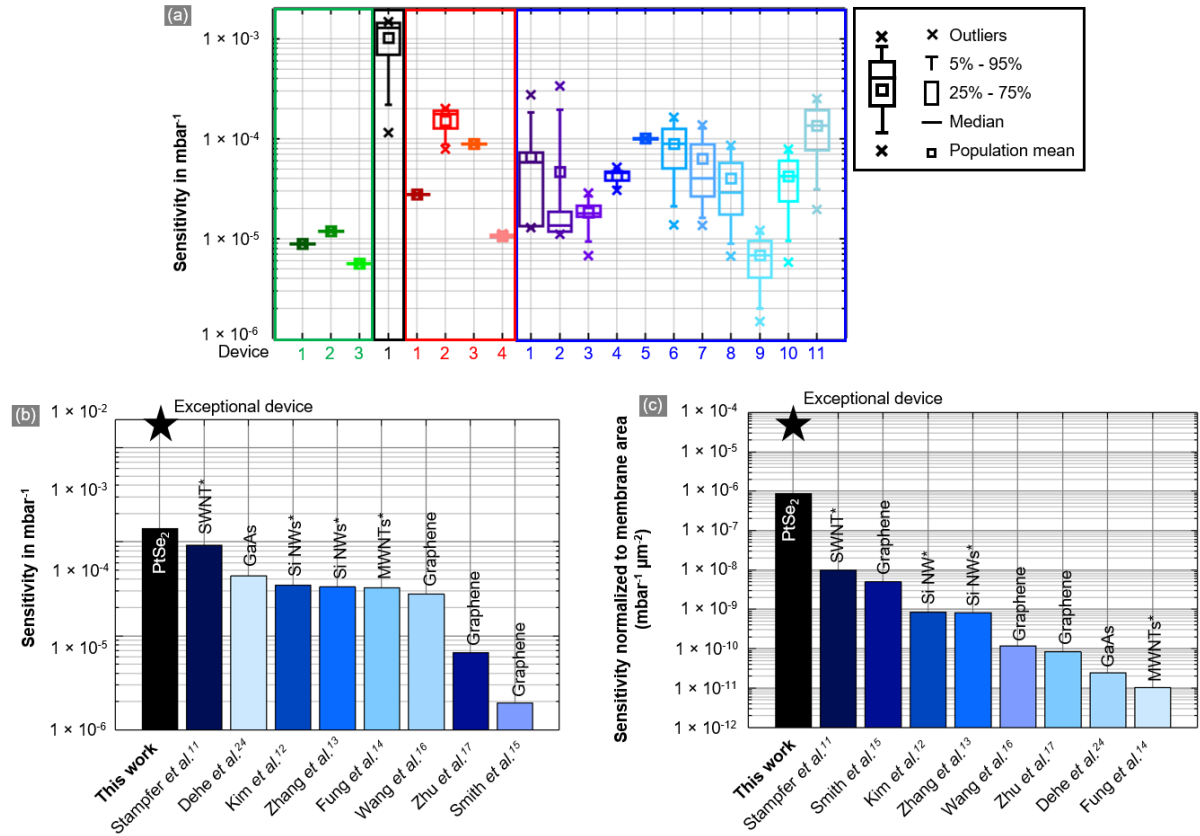

**Figure 3.** Comparison of all measured chips and devices as well as comparison of various pressure sensor devices using different materials for piezoresistive detection. (a) Box and whiskers plot of all measured chips, devices and repeated measurements; (b) Comparison of the sensitivity in mbar-1 with other pressure sensors; (c) comparison of the sensitivity normalized by membrane area in mbar$^{-1}$ $\mu$m$^{-2}$ with other pressure sensors. In plots (b) and (c), devices which require a separate membrane are indicated with a * and the device with exceptionally high sensitivity is indicated by a black star.

A bending beam setup was used to measure another figure of merit, the piezoresistive gauge factor, of the 4.5 nm and 9 nm thick PtS$_2$ films which 7-8 and around 15 layers, respectively. The bending beam setup enables in a very simple way measurements of well-defined strain fields. The gauge factor (GF) can be extracted using equation (2).[28]

$$GF = \frac{\Delta R}{R_0 \cdot \varepsilon} \quad (2)$$

Here, ε is the strain in the bending beam at the position of the sensor (ε = 0.04%). A negative gauge factor (GF) of -84.8 was obtained (Figure 4a, supporting information section S7 and S9). The measured resistance change is shown in Figure 4b and in the supporting information Figure S9b for two PtSe$_2$ layer thicknesses, a mass of 2 kg (Figure 4b) and 0.5 kg (Figure S9c in the supporting information) and multiple cycles with a mass of 2 kg (Figure S9d in the supporting information). Data from a commercially available metal strain gauge is used as a reference (Figure 4b). A decrease in resistance with increasing strain on the PtSe$_2$ film corroborates the integrated pressure sensor measurements (supporting information Figure S9a). The GF of the PtSe$_2$ films exceeds the GF of the metal strain gauge (GF of 2)[28] and of graphene (GF of 2-5)[23,29]. While similar or higher GF have been found for doped silicon (GF of -100 to +200)[28] or mono- and trilayer MoS$_2$ (GF of -148 and -43.5)[30].

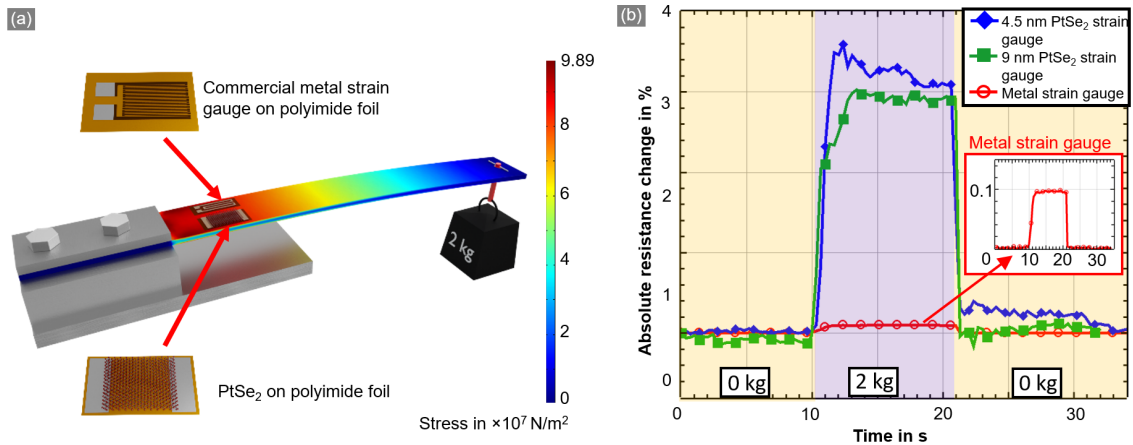

**Figure 4.** (a) Bending beam setup (cantilever beam) with applied PtSe$_2$ and commercially available metal strain gauges including stress simulation with the applied weight; (b) Electrical readout signal during the measurement with an absolute resistance change against time showing two PtSe$_2$ devices (of 4.5 nm PtSe$_2$, diamonds filled in blue and 9 nm PtSe$_2$, squares filled in green) and a metal strain gauge (empty red circles) as reference with an enlarged view in the inset.

DFT calculations were conducted to gain insights into the observed resistance decrease with increased strain in PtSe$_2$ films. The model was fully optimized in terms of the lattice vectors and atomic positions. In the bulk system, which is similar to the experimental conditions in this work, biaxial strain means that the layers are stretched or compressed in plane. As bulk PtSe$_2$ is metallic, the DOS at the Fermi level is found to increase under biaxial tensile strain ($\varepsilon_a$) and reduces under compression, as shown in the supporting information Figure S10a. This result is in accordance with the experimental findings, where the resistivity decreases with increased $\varepsilon_a$.

Next, we have considered the effect of the interlayer strain and compression ($\varepsilon_c$) on the DOS. The results are shown in the supporting information Figure S10b. The compression of the interlayer distance has the opposite influence on the system and it slightly reduces the DOS at the Fermi level. Finally, we have considered the two effects acting in tandem in the following way: for in-plane stretching, there is a compression out of the plane, and vice versa (Figure 5a). The results for the combined effect are shown in Figure 5b. Once again, the DOS at the Fermi level is increased for the in plane stretching and compression in the out-of-plane direction. However, the effect is reduced compared to just $\varepsilon_a$, due to the opposite effect of the interlayer compression, as shown in the supporting information Figure S10b. In comparison to the significant increase of the DOS at the Fermi level upon strain the band structure shows only minor changes (Figure S10c in the supporting information) not indicating strong differences in the effective masses under strain.

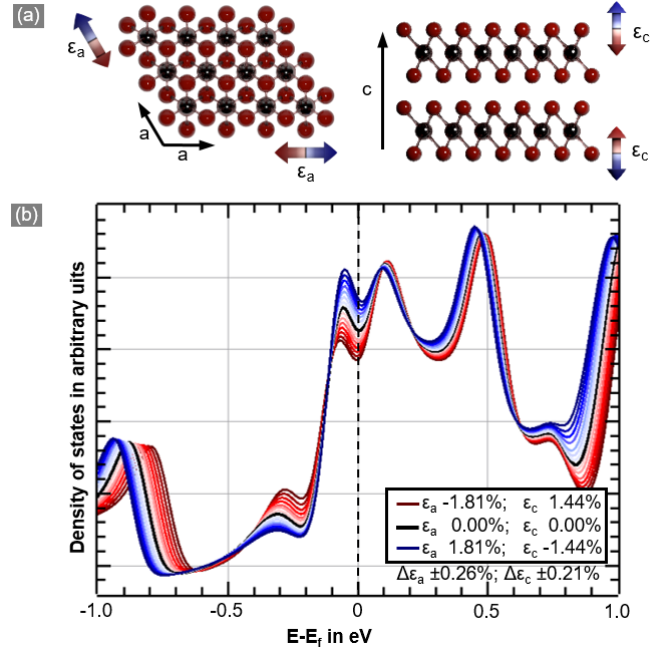

**Figure 5.** (a) Top and side view of the PtSe$_2$ bulk shown with the in-plane, a and out-of-plane, c lattice vectors, and the direction of the applied strain ($\varepsilon_a$ and $\varepsilon_c$). (b) Density of states close to the Fermi level (shifted to zero) under applied tensile strain and compression.

An increase in DOS has also been reported for other single and few layer semiconducting TMDs, like MoS$_2$,[31] MoSe$_2$,[32] WSe$_2$[32] and also PtSe$_2$.[33] The semiconducting materials exhibit a decrease in band gap and even a transition to metallic behavior with applied strain, resulting in a negative gauge factor. For more detailed future studies the influence of many other parameters should be considered and investigated, including the adhesive interface between the steel beam and the polyimide foil, the thickness and mechanical properties of the foil and the PMMA layer covering the PtSe$_2$. Also, the crystallographic orientations, thickness and other material properties of the TAC-grown PtSe$_2$ film need to be investigated and considered in the simulations. Nevertheless, the results of the simulations show a clear indication of an increase in DOS, leading to a decrease in resistance under the applied strain, which is the most plausible cause of the experimentally extracted negative GF.

We have demonstrated $PtSe_2$-based piezoresistive pressure sensors with high gauge factors and very high sensitivity. The experiments were conducted on large-area $PtSe_2$ thin films grown by a scalable method of thermal conversion of metallic films at a low temperature of 400°C. A negative gauge factor of approximately -85 was obtained from a $PtSe_2$ strain gauge through bending beam experiments and can be understood in terms of a strain-induced increase in the density of states (DOS), as verified through DFT simulations. Furthermore, integrated piezoresistive pressure sensors were fabricated using $PtSe_2$/PMMA membranes. The $PtSe_2$ pressure sensors showed very high sensitivity, outperforming piezoresistive pressure sensors based on other materials by at least a factor of 82. When normalized to the active membrane area, the sensitivity of our devices is almost two to five orders of magnitude larger than reported values from any other technology till date. One exceptional device has an even higher sensitivity of $1.64 \times 10^{-3}$ $mbar^{-1}$ and outperforms existing technologies by an even greater margin. These results suggest that layered 2D $PtSe_2$ has great potential for future piezoresistive NEMS applications, beyond the demonstrated pressure sensors. These devices have a footprint that is up to four orders of magnitude smaller than other pressure sensors while showing a similar or even higher sensitivity. Further down-scaling to smaller cavities will improve the stability and yield of the membranes, and allow decreasing the membrane thickness to maintain the high sensitivity. The low growth temperature and general CMOS material compatibility makes $PtSe_2$ highly attractive for future silicon technology integration.

## Methods

**Materials Synthesis and analysis**

Pt layers with a thickness of 1 and 2 nm were sputter deposited on 1.5 x 1.5 cm$^2$ Si/SiO$_2$ substrates using a Gatan coating system (Gatan 682 PECS) with a 19 mm diameter Pt target. A TAC process was used to selenize the pre-deposited Pt layers, as described in detail in references.[4,7] The cross-sectional Raman measurements were conducted using a WITec alpha 300R system with a 532 nm laser as source of excitation. A laser power of 75 $\mu$W was used. For obtaining single Raman spectra an 1800 g/mm grating was used, while the depth scan (cross-section) was conducted with a 300 g/mm grating. The system is limited to a lateral and vertical resolution of 300 nm and 900 nm, respectively.

**Strain gauge fabrication**

The strain gauges were fabricated by transferring as-grown PtSe$_2$ films onto pre-patterned polyimide foil substrates. A typical polymer-supported transfer method was used for the film transfer. 200 nm thick PMMA (Allresist AR-P 649.04) was spun on the as-grown PtSe$_2$ samples at a speed of 2000 rpm for 50 s. After spin-coating, the samples were annealed for 5 min at 80°C to cure the PMMA. 2 molar KOH was used to etch away the SiO$_2$ layer and release the PMMA-covered PtSe$_2$ films from the substrate. The PMMA-covered PtSe$_2$ films were cleaned in DI-water, followed by a transfer to the pre-patterned (500 nm thick copper contacts from evaporation) polyimide foil. After drying, the foil was glued (super glue Z70 by HBM) onto the steel beam (size of 300 × 30 mm$^2$, 3 mm thick), 200 mm away from the loading point. A commercially available metal strain gauge was glued next to the PtSe$_2$ strain gauge. Cables were then soldered onto the contact pads for electrical connections.

**Pressure sensor fabrication**

A Si chip (7 × 7 mm$^2$) with 1.6 $\mu$m thick SiO$_2$ was used as a device substrate (Figure S2a in the supporting information). The cavities have circular geometry (radii between 2 $\mu$m and 20 $\mu$m) with arrays of 1 to 15 cavities and rectangular geometries (length of 20 $\mu$m and width between 2 $\mu$m and 4 $\mu$m) in arrays between 1 and 3 cavities. They were structured by photolithography, followed by Ar- and CHF$_3$-based reactive ion etching (RIE, 200 mW, 40 mTorr) to create a vertical profile with 1.4 $\mu$m of depth (Figure S2b in the supporting information). The contact regions were defined by photolithography, followed by a RIE process to embed the contacts and self-aligned metal evaporation (Cr/Au, 40 nm/400 nm, Figure S2c in the supporting information). The PtSe$_2$ films grown from 1 nm initial Pt films were transferred onto the device substrate using a PMMA-supported transport method as described above (Figure S2d in the supporting information). Lastly, the chip was wire bonded into a 44-pin PLCC chip package using 25 $\mu$m diameter Au wire.

**Simulation**

All calculations were performed using the DFT method as implemented in the VASP plane wave code.[34,35] Ion-electron interactions were described by projector augmented wave approximation.[36] We have employed Perdew–Burke–Ernzerhof (PBE)[37] functional under the generalized gradient approximation. Van der Waals interaction corrections were included using the D3 approach proposed by Grimme.[38] Cutoff energy was set to 600 eV and the convergence threshold for residual force was 0.01 eV Å$^{-1}$. Brillouin zone integration was carried out at 6×6×6 Monkhorst–Pack k-grids. Spin-orbit coupling (SOC) was included in the electronic structure calculations. Bulk PtSe$_2$ was fully optimized (lattice vectors and atomic positions) resulting in the following lattice vectors: a = b = 3.788 Å and c = 4.790 Å, and a distance between Se atoms in the Se-Pt-Se sandwich of 2.550 Å. This is in good agreement with the experimental data by Wang et al.,[6] with a = b = 3.70 Å and the distance between Se atoms of 2.53 Å.

SUPPORTING INFORMATION

A schematic of the growth process for PtSe$_2$ is shown with additional Raman spectra and a TEM cross-section of the layered films. The fabrication process is described in a schematic and the measurement chamber is shown in a schematic. I-V characteristics of the unstrained pressure sensor are shown as well as additional electrical characterization data of the pressure sensors. A table compares pressure sensors from literature. The strain gauge measurements are described in detail, including the I-V characteristics of the devices as well as further electrical data. Further DFT calculations are shown for bulk and for three layer PtSe$_2$.

AUTHOR INFORMATION

**Corresponding Author**

*Tel: +49 241 8867 200, Email: max.lemme@eld.rwth-aachen.de; lemme@amo.de

**Author Contributions**

The manuscript was written through contributions of all authors. All authors have given approval to the final version of the manuscript.

ACKNOWLEDGMENT

Funding from the M-ERANET/German Federal Ministry of Education and Research (BMBF, NanoGraM, 03XP0006), the European Research Council (ERC, InteGraDe, 3017311), FLAG-ERA (HE 3543/27-1), European Union Seventh Framework Program: Graphene Flagship (Grant No. 649953), Science Foundation Ireland (PI Grant No. 15/IA/3131, 12/RC/2278 and 15/SIRG/3329) and the German Research Foundation (DFG LE 2440/1-2 and HE 3543/18-1) are gratefully acknowledged.

The authors would like to thank Anderson Smith for etching the cavities of the chips and for fruitful discussions and Martin Otto for conducting the AFM measurements. The ZIH Dresden is gratefully acknowledged for the computer time.

ABBREVIATIONS

CCR2, CC chemokine receptor 2; CCL2, CC chemokine ligand 2; CCR5, CC chemokine receptor 5; TLC, thin layer chromatography.

# Supporting information

# Highly sensitive electromechanical piezoresistive pressure sensors based on large-area layered PtSe$_2$ films

Stefan Wagner[1], Chanyoung Yim[2], Niall McEvoy[3], Satender Kataria[1], Volkan Yokaribas[4], Agnieszka Kuc[5], Stephan Pindl[6], Claus-Peter Fritzen[4], Thomas Heine[5], Georg S. Duesberg[2,3], Max C. Lemme[1,2,7*]

[1]Chair of Electronic Devices, Faculty of Electrical Engineering and Information Technology, RWTH Aachen University, Otto-Blumenthal-Str. 2, 52074 Aachen, Germany

[2]Institute of Physics, EIT 2, Faculty of Electrical Engineering and Information Technology, Universität der Bundeswehr München, Werner-Heisenberg-Weg 39, 85577 Neubiberg, Germany

[3]Centre for Research on Adaptive Nanostructures and Nanodevices (CRANN), Advanced Materials and BioEngineering Research (AMBER) and School of Chemistry, Trinity College Dublin, Dublin 2, Ireland

[4]Department of Mechanical Engineering, University of Siegen, 57076 Siegen, Germany

[5]Wilhelm-Ostwald-Institute für Physikalische und Theoretische Chemie, Universität Leipzig, Linné Str. 2, 04103 Leipzig, Germany

[6]Infineon Technologies AG, Wernerwerkstraße 2, 93049 Regensburg, Germany

[7]Advanced Microelectronic Center Aachen (AMICA), AMO GmbH, Otto-Blumenthal-Str. 2, 52074 Aachen, Germany

*Tel: +49 241 8867 200, Email: max.lemme@rwth-aachen.de; lemme@amo.de

# S1: PtSe$_2$ synthesis via a TAC process

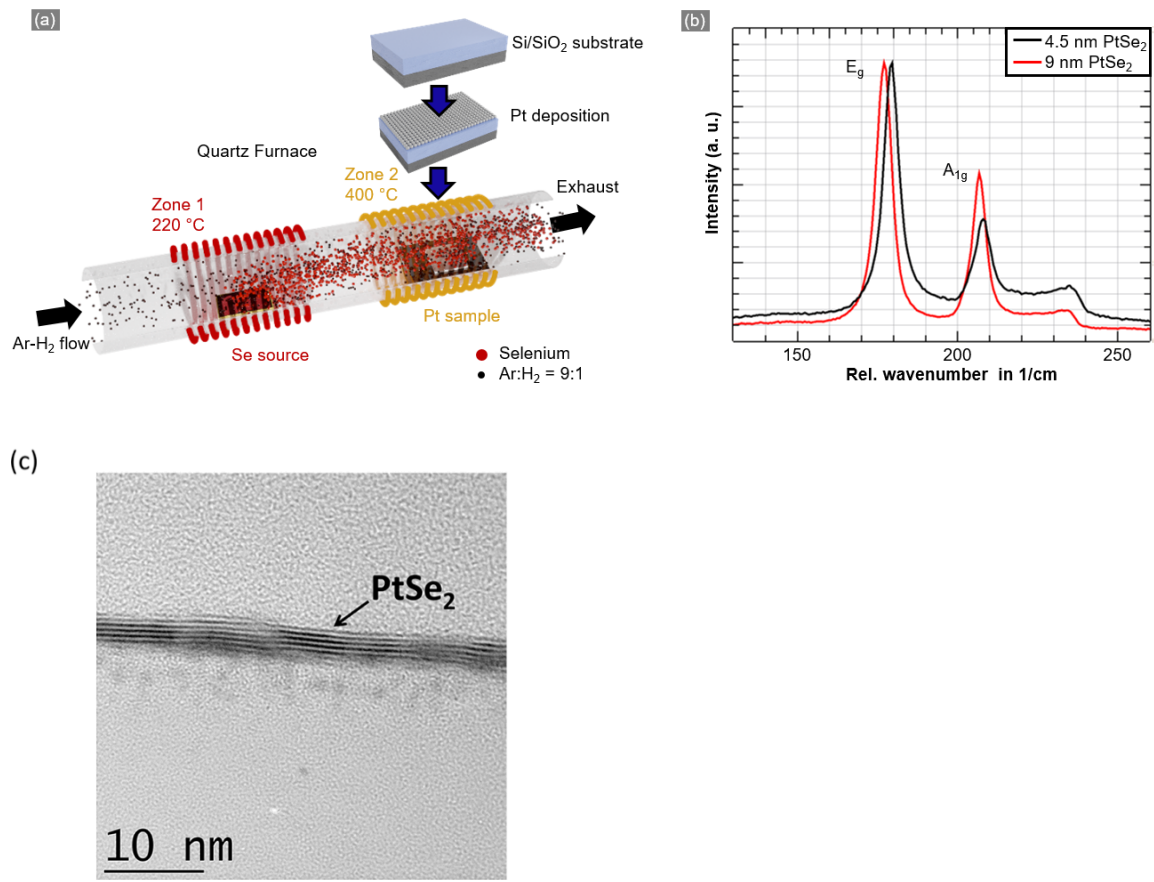

*Figure S5: (a) Pt is deposited on a Si/SiO$_2$ substrate and placed into a quartz tube furnace. The Se source is evaporated in zone 1 at 220 °C and transported by the Ar-H$_2$ flow to zone 2, selenizing the Pt sample at 400 °C. (b) Raman spectra of 4.5 nm and 9 nm PtSe$_2$ films. (c) TEM cross-sectional image of a layered PtSe$_2$ film grown by TAC (adapted from [1]).*

## S2: Fabrication process of the pressure sensor

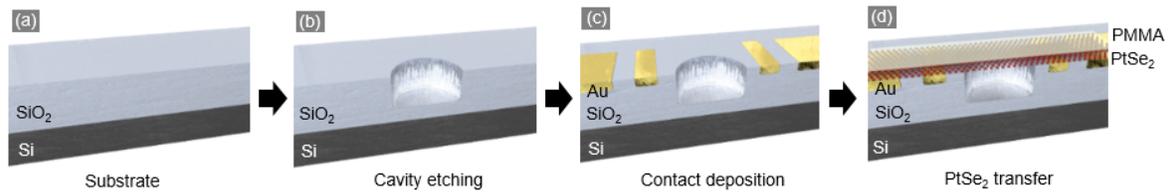

*Figure S6: (a)-(d) A summary of the device fabrication process; (a) starting with a Si/SiO$_2$ substrate, (b) reactive ion etched (RIE) cavity, (c) the embedded metal contact deposition and (d) the transfer of PtSe$_2$ covered by a PMMA layer. See Methods for the detailed fabrication process description.*

**S3: Pressure chamber schematic**

The chamber is built using standard vacuum components. Sensors included in the chamber are the humidity sensor HIH-4000 (Honeywell International Inc.), the temperature LM35 (Texas Instruments) sensor and the pressure sensor MXP2200AP (NXP Semiconductors).

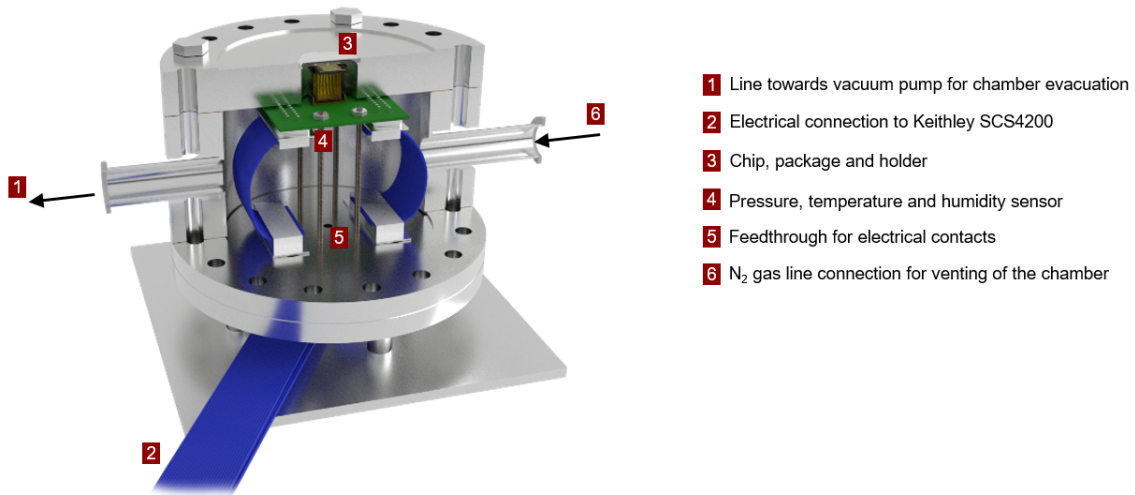

1. Line towards vacuum pump for chamber evacuation
2. Electrical connection to Keithley SCS4200
3. Chip, package and holder
4. Pressure, temperature and humidity sensor
5. Feedthrough for electrical contacts
6. $N_2$ gas line connection for venting of the chamber

*Figure S3: Home-built vacuum chamber design for the electrical measurement of NEMS pressure sensors.*

## S4: I-V curve at environmental pressure (unstrained)

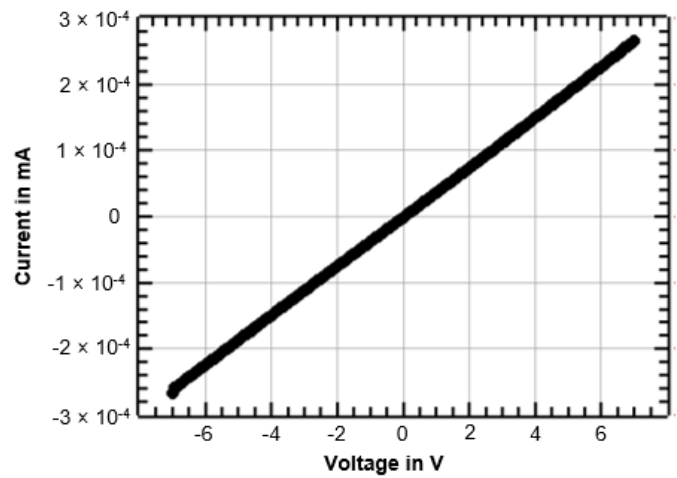

*Figure S4: I-V curve of the PtSe$_2$ channel (1 nm initial Pt thickness) used as membrane for the pressure sensor at environmental pressure.*

## S5: Further measurements of the PtSe$_2$ pressure sensor

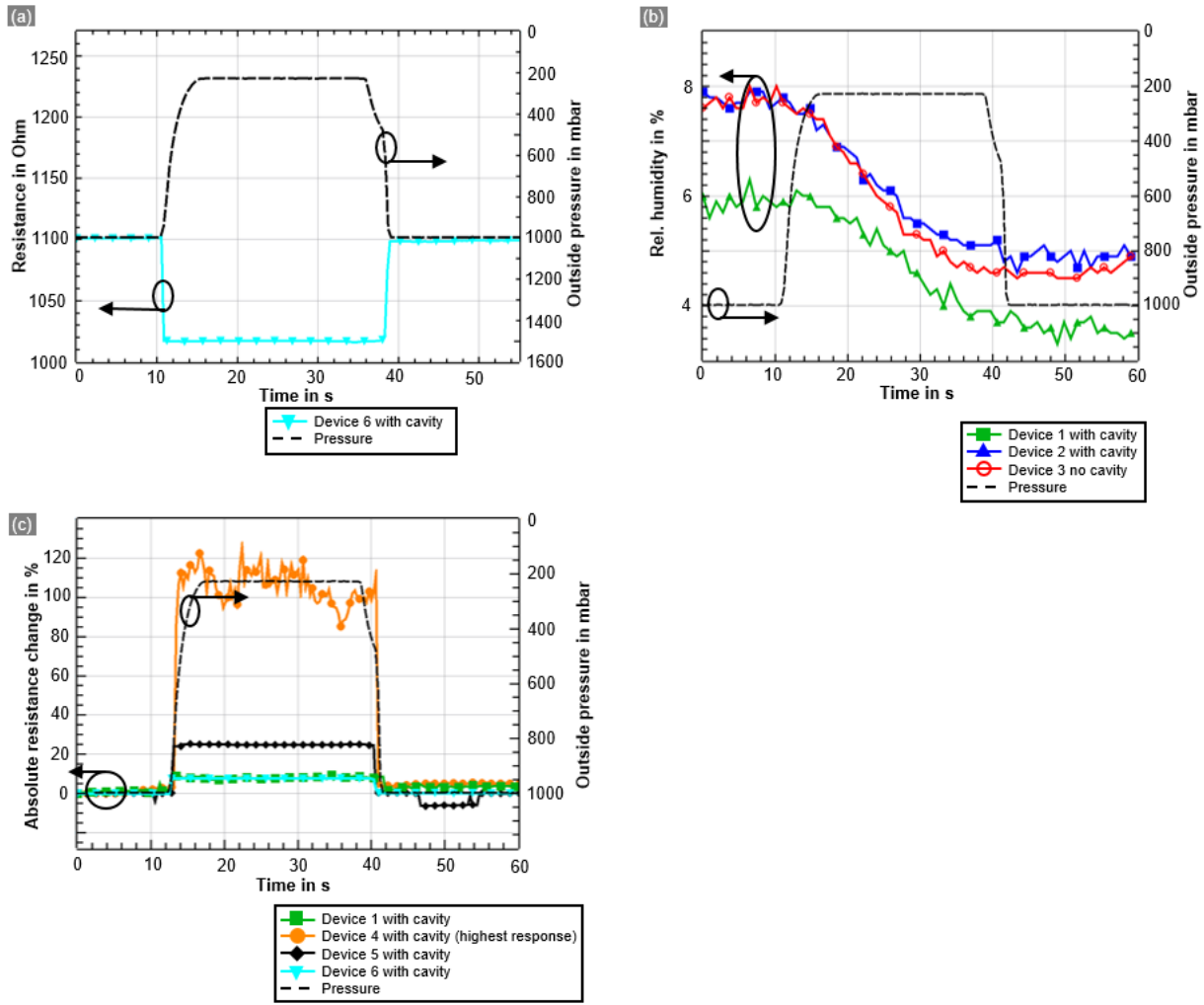

*Figure S5: (a) Measured values plotted as resistance in Ohm against the time with a resistance decrease under strain with decreasing pressure of the PtSe$_2$ pressure sensor (triangles filled in turquoise); (b) Relative humidity of three devices (two devices with and one device without cavity as a reference) against the time with indicated pressure condition during measurement; (c) More measured devices with the resistance change in % (left y-axis) against the measurement time and the pressure during the experiment (right y-axis): device 1 for comparison, device 4 with the highest response detected and device 5 and 6 measured on a new chip with PtSe$_2$ from 1 nm starting Pt.*

# S6: Comparison of various pressure sensor devices

*Table S1: Comparison of various pressure sensor devices ordered by the normalized sensitivity by membrane area.*

| Device structure | Membrane area ($\mu m^2$) | Sensitivity (mbar$^{-1}$) | Sensitivity per membrane area (mbar$^{-1}$ $\mu m^{-2}$) | Publication year | References |
|---|---|---|---|---|---|
| Suspended PtSe2 covered with PMMA (exceptional device) | 5.00E+01 | 1.64E-03 | 3.28E-05 | 2017 | This work (highest value) |
| Suspended PtSe$_2$ covered with PMMA | 1.57E+02 | 1.39E-04 | 8.85E-07 | 2017 | This work |
| SWNT | 9.16E+03 | 9.26E-05 | 1.01E-08 | 2006 | Stampfer et al.[2] |
| Suspended Graphene | 3.84E+02 | 1.94E-06 | 5.05E-09 | 2013 | Smith et al.[3] |
| Si nanowires | 4.00E+04 | 3.47E-05 | 8.67E-10 | 2009 | Kim et al.[4] |
| Si nanowires | 4.00E+04 | 3.33E-05 | 8.33E-10 | 2016 | Zhang et al.[5] |
| Suspended graphene on a perforated SiNx membrane | 2.40E+05 | 2.80E-05 | 1.17E-10 | 2016 | Wang et al.[6] |
| Graphene on a suspended imperforated SiNx membrane | 7.84E+04 | 6.67E-06 | 8.50E-11 | 2013 | Zhu et al.[7] |
| GaAs/AlGaAs | 1.77E+06 | 4.37E-05 | 2.47E-11 | 1995 | Dehe et al.[8] |
| MWNT embedded into a PMMA diaphragm | 3.14E+06 | 3.27E-05 | 1.04E-11 | 2005 | Fung et al.[9] |

**S7: PtSe₂ strain gauge**

The PtSe₂ films were transferred onto a polyimide foil with prefabricated metal contacts to form a PtSe₂ strain gauge. The PMMA layer was left on the device to encapsulate the device and protect it against environmental influences, like humidity to avoid cross-sensitivity of the sensor and decrease the noise during the measurement. The strain gauge was then placed on a steel beam with known mechanical properties and glued at a position where high mechanical stress was expected (200 mm away from the point of load, main manuscript, Fig. 4a). Simulations of the stress distribution in the beam under load yield strain of about 0.04% at the sensing position, i.e. where PtSe₂ is fixed.[10] With dimensions of the free beam (300 mm x 30 mm x 3 mm) and the distance between the sensing position and suspension point of the weight (200 mm), the stress (σ) which a sensor experiences, can be calculated from equation (S7.1, S7.2), where *WT* is the weight attached to the beam, $x_m$ is the distance from the sensor to the suspension point, *w* is the beam width, *t* is the beam thickness, and *E* is the Young's modulus of the steel beam (0.2 TPa).

$$\sigma = \frac{WT \cdot x_m \cdot 6}{w \cdot t^2} \tag{S7.1}$$

$$\varepsilon = \frac{\sigma}{E} \tag{S7.2}$$

The calculated stress at the sensor is estimated to 8.71 x 10⁷ N/m², therefore, a strain (ε) of 4.36 × 10⁻⁴ is extracted, which agrees well with the simulations.

A commercially available metal strain gauge was placed next to the PtSe₂ device as a reference. While a static bending beam experiment was conducted, the total resistance of the gauges was recorded at +1 V of dc bias. The current-voltage (I-V) characteristics of the PtSe₂ device for the unloaded case shows good linearity, i.e. the transferred PtSe₂ forms good Ohmic contacts with metal (copper) electrodes (supporting information, Fig. S8). The electrical response of the devices was measured during loading and unloading of a mass of 2 kg attached at the end of the cantilever beam. Fig. 4b of the main manuscript compares the absolute resistance change in percentage (%) for the PtSe₂ devices to a commercially available metal strain gauge over the measurement time. The applied strain causes a resistance change in the PtSe₂ films, implying the presence of a piezoresistive effect. The gauge factor has been calculated using equation (S7.3), where *ΔR* is the difference in resistance, *R* is the initial resistance in the unloaded case and ε is the strain.

$$GF = \frac{\Delta R}{R \cdot \varepsilon} \qquad (S7.3)$$

The resistance of the PtSe$_2$ gauge decreases with applied strain, in contrast to a resistance increase of the metal strain gauge (supporting information, Fig. S9a). The electrical data result in an average negative GF of -84.8 for the PtSe$_2$ sensor. The GF of the commercial metal strain gauge is 2.1, in agreement with the specifications.[10] The measurements were conducted for two different PtSe$_2$ thicknesses of 4.5 nm and 9 nm and repeated several times (supporting information, Fig. S9b). The measured devices highlight the consistent performance of the PtSe$_2$ film. The GF of these PtSe$_2$ layers is approximately 40 times higher than in metal [10] and graphene,[3,11] and comparable to MoS$_2$-based strain gauges.[12,13] Due to its 2D nature, the PtSe$_2$-based strain gauge is believed to be more advantageous in flexible electronics than metal strain gauges suggesting high potential for applications in the area of wearable electronics.

## S8: I-V curve of an unloaded strain gauge

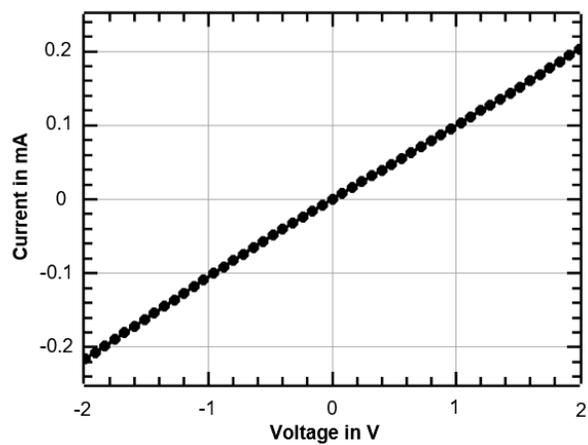

*Figure S8: I-V curve of the PtSe$_2$ (4.5 nm thick PtSe$_2$) strain gauge for the unloaded case.*

# S9: Further measurements of the PtSe₂ strain sensor

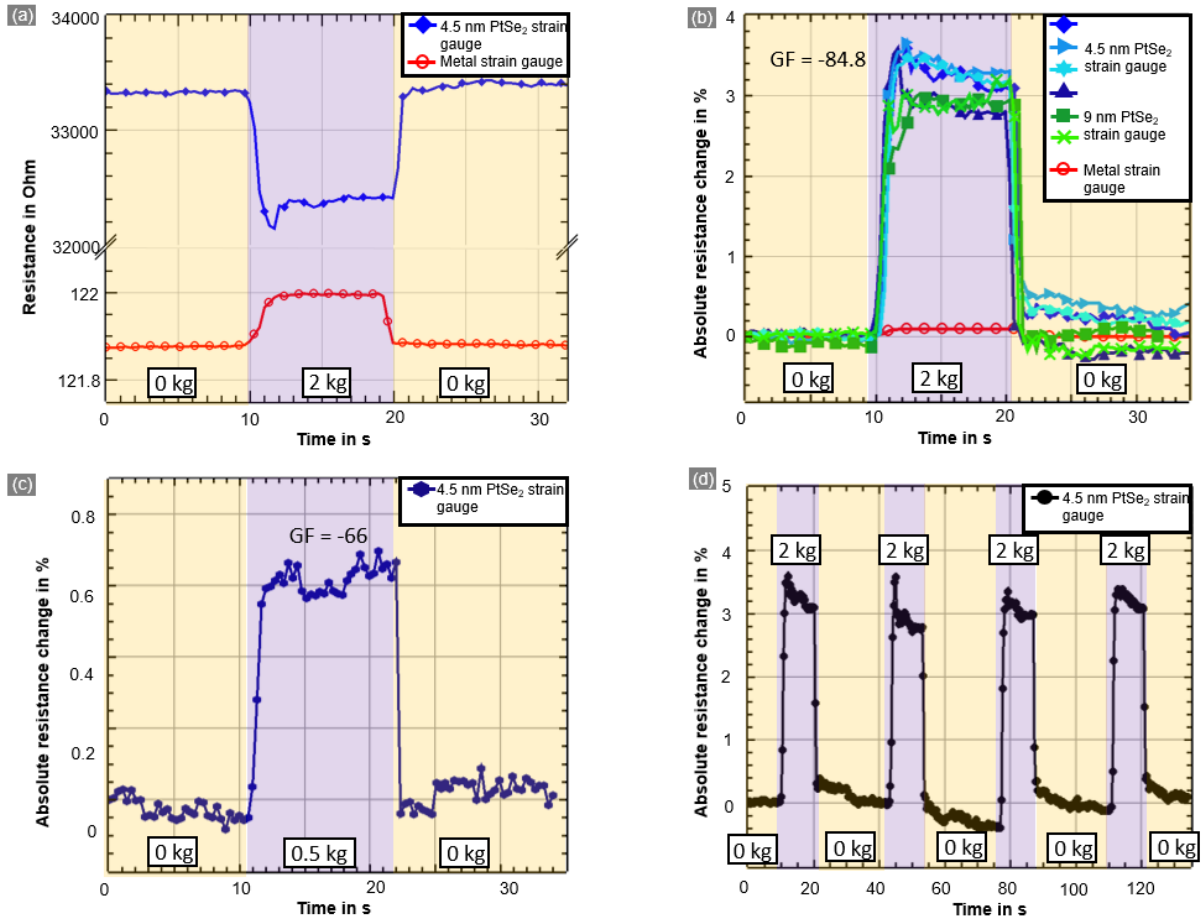

*Figure S9: a) Measured values plotted as resistance in Ohm against the time with the metal strain gauge (empty circles with red outline), with a resistance increase under load and the PtSe₂ strain gauge (diamond filled in blue) with a resistance decrease under load; b) Plot of multiple measurements of the PtSe₂ strain gauges with PtSe₂ of 4.5 nm (shades of blue color) and 9 nm (shades of green color) thickness as well as the reference metal strain gauge (empty circles with red outline); (c) Bending beam measurement for a 4.5 nm thick PtSe₂ film and a mass of 0.5 kg; (d) Multiple cycles with a mass of 2 kg for a 4.5 nm thick PtSe₂ film.*

## S10: DFT calculations

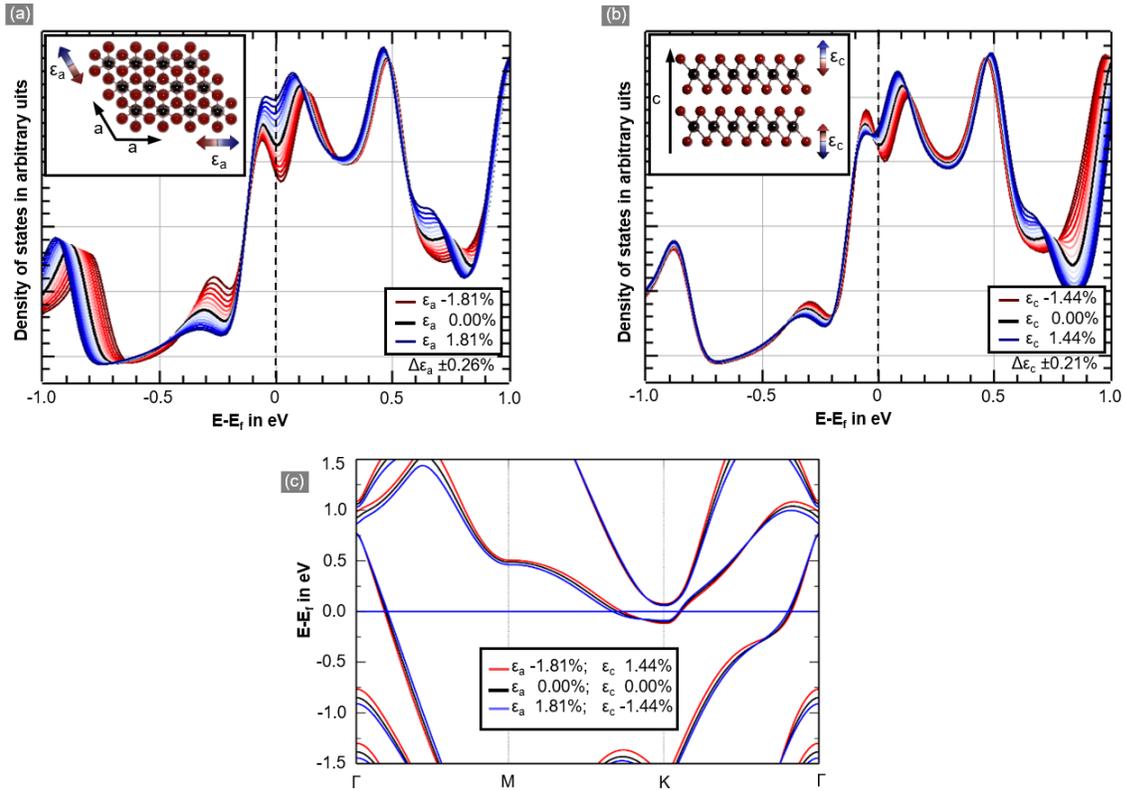

*Figure S10: (a) Density of states close to the Fermi level (shifted to zero) under applied tensile strain and compression. The inset shows a top view of the PtSe$_2$ bulk with the in-plane lattice vector a and the direction of the biaxial strain ($\varepsilon_a$). (b) Density of states close to the Fermi level (shifted to zero) under applied tensile strain and compression. The inset shows the side view of the PtSe$_2$ bulk shown with the out-of-plane lattice vector c and the direction of the interlayer strain ($\varepsilon c$); (c) the band structure of bulk PtSe$_2$.*

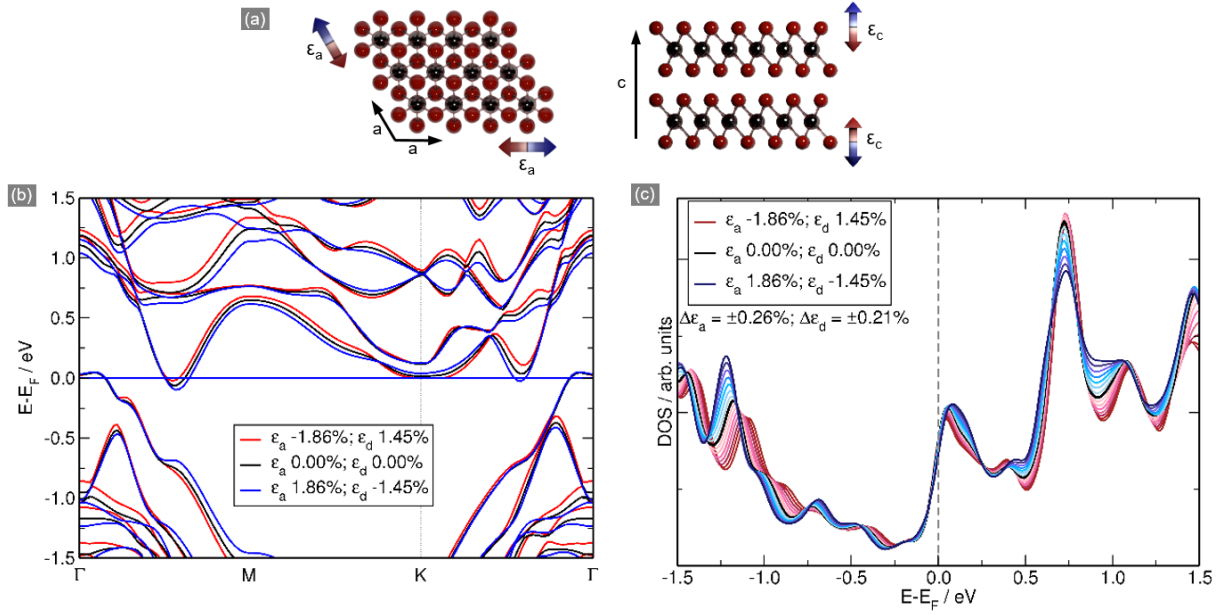

*Figure S11: (a) Top and side view of the 3 layers PtSe$_2$ shown with the in-plane a and out-of-plane c lattice vectors, and the direction of the applied strain ($\varepsilon_a$ and $\varepsilon_c$). (b) Band structure plot of 3 layers PtSe$_2$ under applied tensile and compressive strain; (c) Density of states close to the Fermi level (shifted to zero) under applied tensile strain and compression for 3 layers PtSe$_2$.*